\documentclass{elsart}

\usepackage{natbib}

\usepackage{graphicx}

\usepackage{amssymb}

\begin{document}

\begin{frontmatter}



\title{SILEM : a new gaseous detector with integrated $x-y$ readout plane}


\author{Patrick Weber \ead{patrick.weber@unine.ch}}

\address{Physics Department, Neuch\^atel University, A-L. Breguet 1, CH-2000 Neuch\^atel, Switzerland}


\begin{abstract}
This works reports on gaseous detectors developments made in the frame of the EXO double-beta decay experiment. LEM (Large Electron Multiplication) are electron amplification grids based on GEM. They were developed in Neuch\^atel \cite{jeanneret} and showed remarquable resistance to sparks. The new SILEM grid combines  the properties of the standard LEM with a micropatterned $x-y$ readout plane  on one of the grid side. It allows thus the amplification of the primary electrons and their position in the $x-y$ plane. 

\end{abstract}

\begin{keyword}
Gaseous detector \sep Time Projection Chamber


\end{keyword}

\end{frontmatter}


\section{Introduction}
\label{}

When used in TPC (Time Projection Chamber), hole-type gaseous detectors such as the GEM \cite{GEM} are able to amplifiy the signal of primary electrons --- due to ionization of the incident particle while traversing the TPC volume ---  with gains up to $10^{4}$. Electron amplification gives informations on the incident particles energy, but also allow a detection of the electrons positions, when a readout plane is placed behind it. 
However, GEM are very sensitive to breakdowns. Their structure can be broken when sparks occur. \\
LEM gaseous detectors were developed at Neuch\^atel \cite{jeanneret}; these amplification grids are machined in a printed board circuit substrate with a uniform copper film on both sides. The holes are mechanically drilled. LEM holes diameter are $500\mu m$ and LEM thickness 1mm. LEM exhibit gains up to $10^{3}-10^{4}$. Their simplicity of machining and resistance to sparks made them an efficient alternative to GEMs. \\
The SILEM grid is an optimization of the LEM geometry in order to obtain the best possible gain. Moreover, an $x-y$ pixels micropatterned readout plane on one side of the grid is integrated and allows to measure the position of the primary electrons. The SILEM thus combines the amplification system and the readout plane.  


\section{Simulation}

\begin{figure} [h!]
\begin{centering}
\includegraphics[scale=0.35]{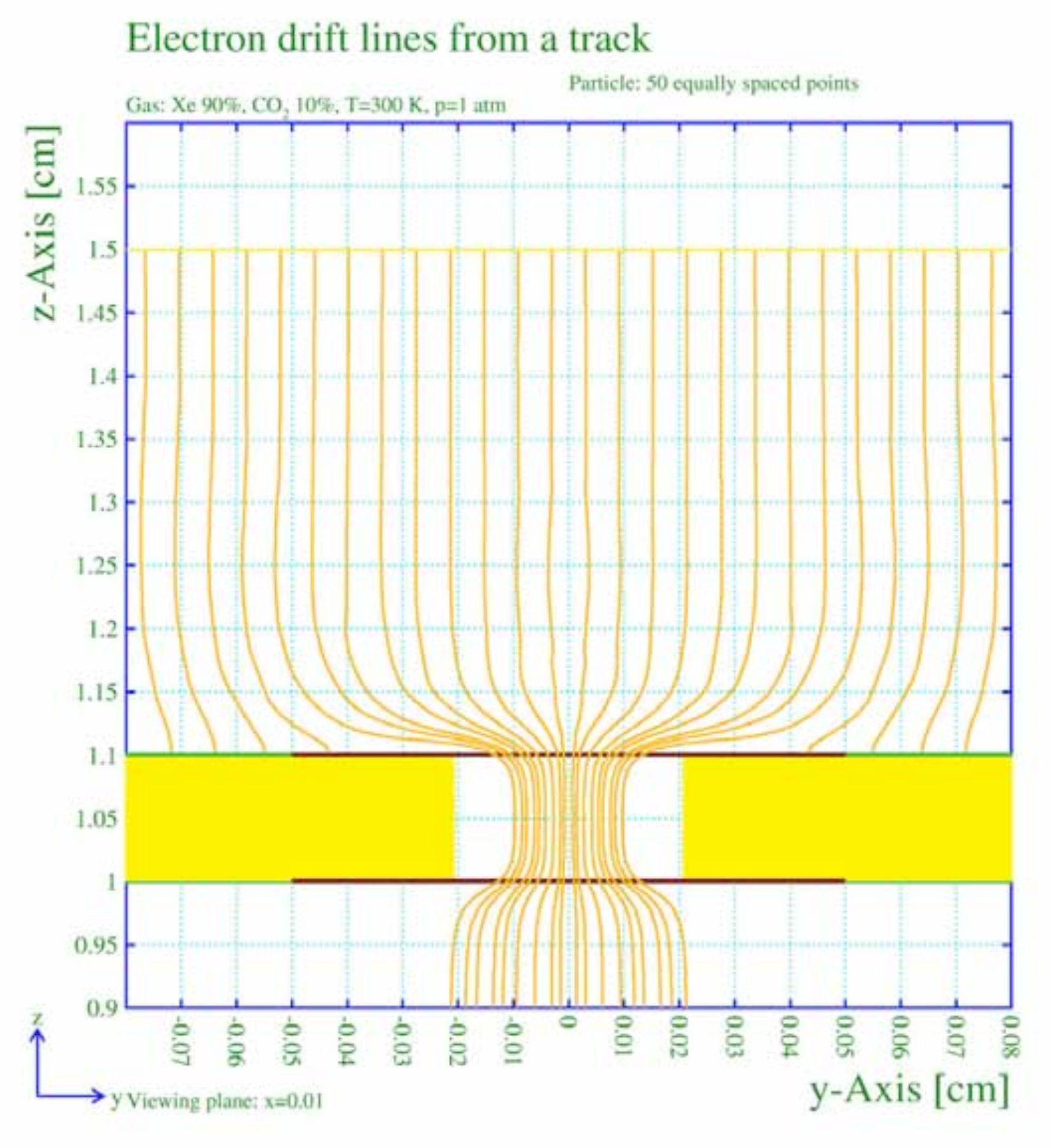}
\caption{Garfield simulation of LEM field lines.}
\label{LEM}
\end{centering}
\end{figure}

The LEM amplification grid had a non-optimized geometry in the sense that some electric field lines did not pass through the LEM holes, as shown on Figure \ref{LEM}. Consequently, $\sim20\%$ of the primary electrons were lost on the LEM anode and were not be amplified. 

A Maxwell/Garfield simulation \cite{maxwell}, \cite{garfield} with various LEM geometries has been performed, leading to the conclusion that the ratio LEM thickness / holes diameter has to be in the range:
$$1\leq \frac{t}{d}\leq2$$

Large holes allow to avoid an ion screening effect, i.e. electronic shower in the LEM holes ionizes the gas and thus creates ions, drifted "upwards" due to the electric field. An important density of ions in a hole obstructs the primary electrons entering the holes. Moreover, the LEM gain is proportional to the hole length --- i.e. to the LEM thickness --- as long holes allow the development of an electromagnetic shower. However, too long holes stops the secondary electrons produced in the holes and inhibit the electrons readout after the amplification grid.
The SILEM grid has $300 \mu m$ holes and is $420 \mu m$ thick. Its $t/d$ ratio is then 1.4. \\

\begin{figure}[h]
\begin{center}
\begin{tabular}{cc} 
\includegraphics[scale=0.32]{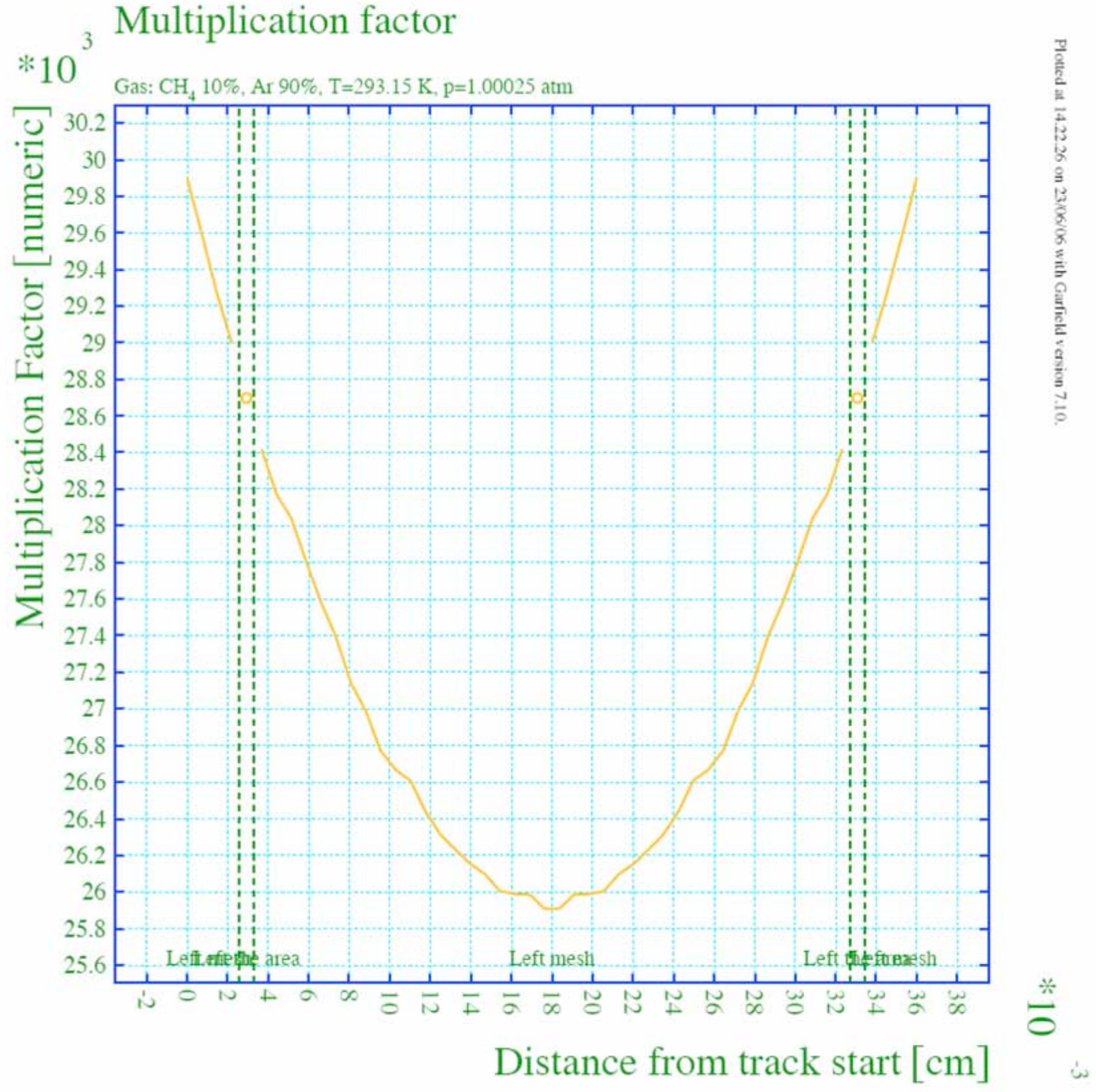} & 
\includegraphics[scale=0.32]{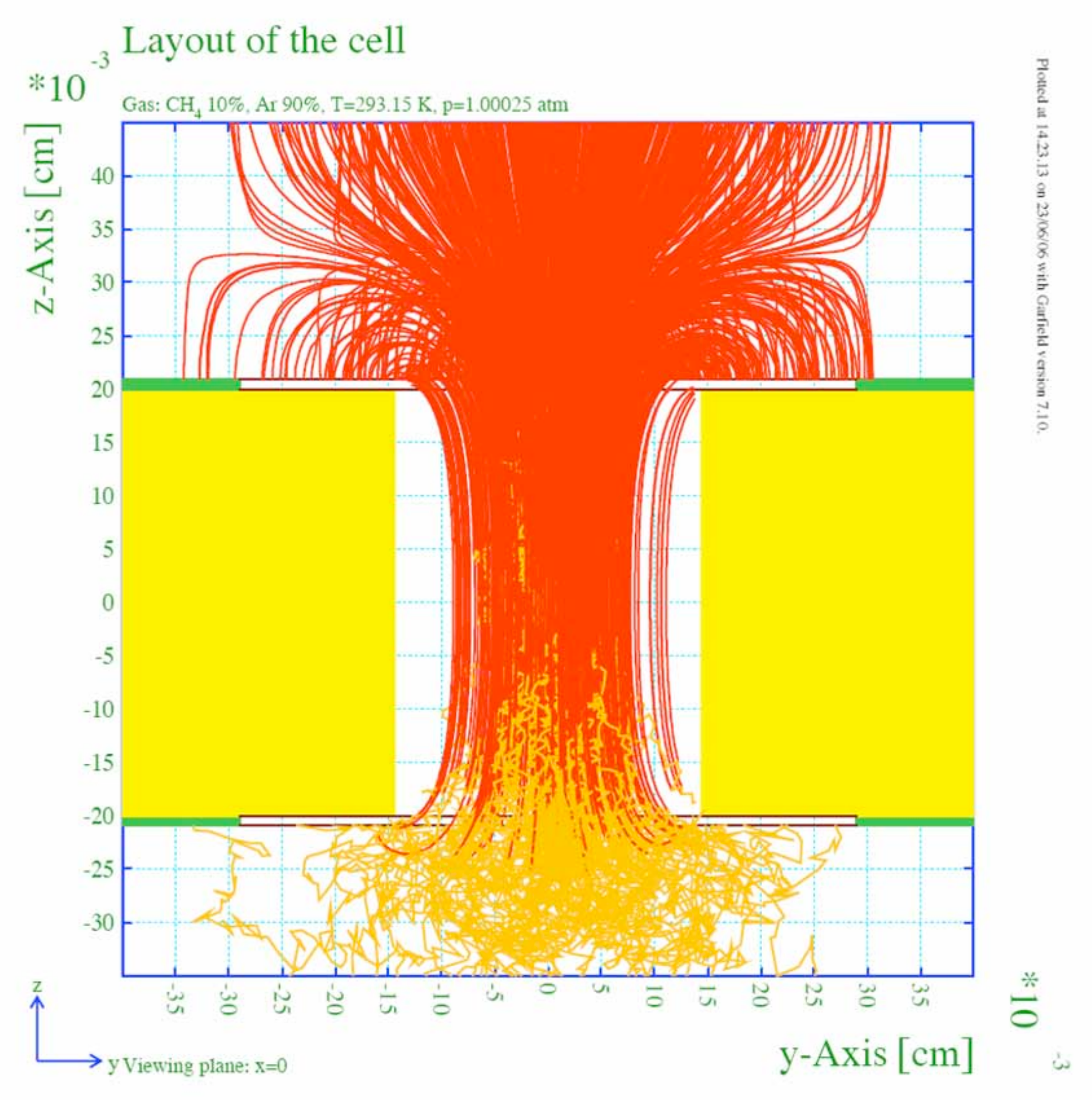} 
\end{tabular}
\end{center}
\caption{SILEM gain and avalanche. Garfield simulation.}
\label{avalanche}
\end{figure}

The Maxwell/Garfield simulation allows to simulate the grid gain and avalanche, as shown on Figure \ref{avalanche}.

As the gain is not constant across the hole, this parabola shape could be used to increase the spatial resolution of the grid. The $15-20\%$ variation of the gain on the hole section would permit to give an additional information on the electron position, if the energy resolution is smaller compared to this gain variation.

\begin{figure} 
\begin{centering}
\includegraphics[scale=0.4]{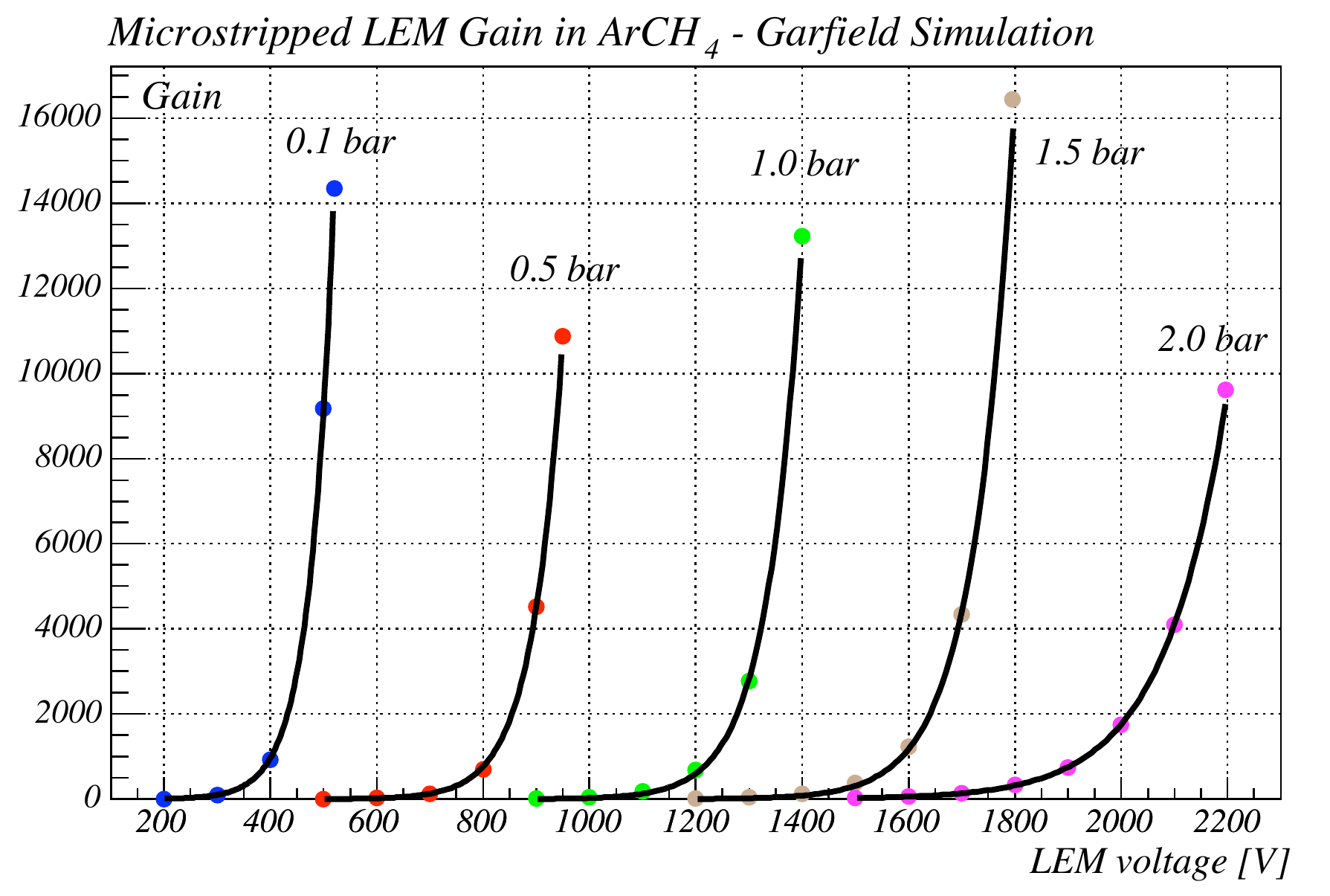}
\caption{SILEM gain in P10 gas. Garfield simulation.}
\label{gain}
\end{centering}
\end{figure}

Figure \ref{gain} is the simulation result for the SILEM gain in various $ArCH_{4}$ (P10 - 90\% $Ar$, 10\% $CH_{4}$) pressures. However, breakdowns are not simulated.

\section{SILEM design and fabrication}

\begin{figure}[h]
\begin{center}
\begin{tabular}{cc} 
\includegraphics[scale=0.4]{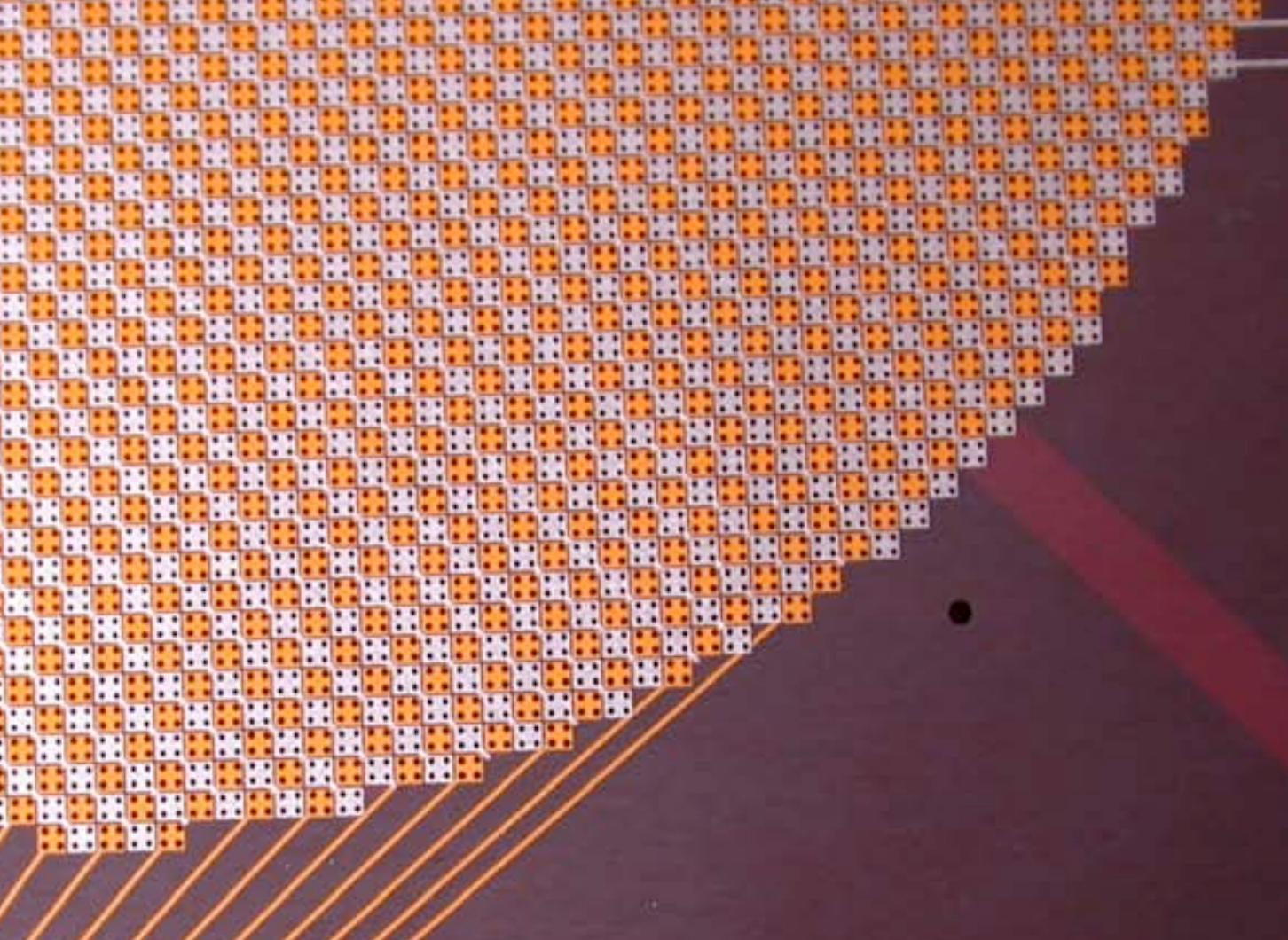} & 
\includegraphics[scale=0.22]{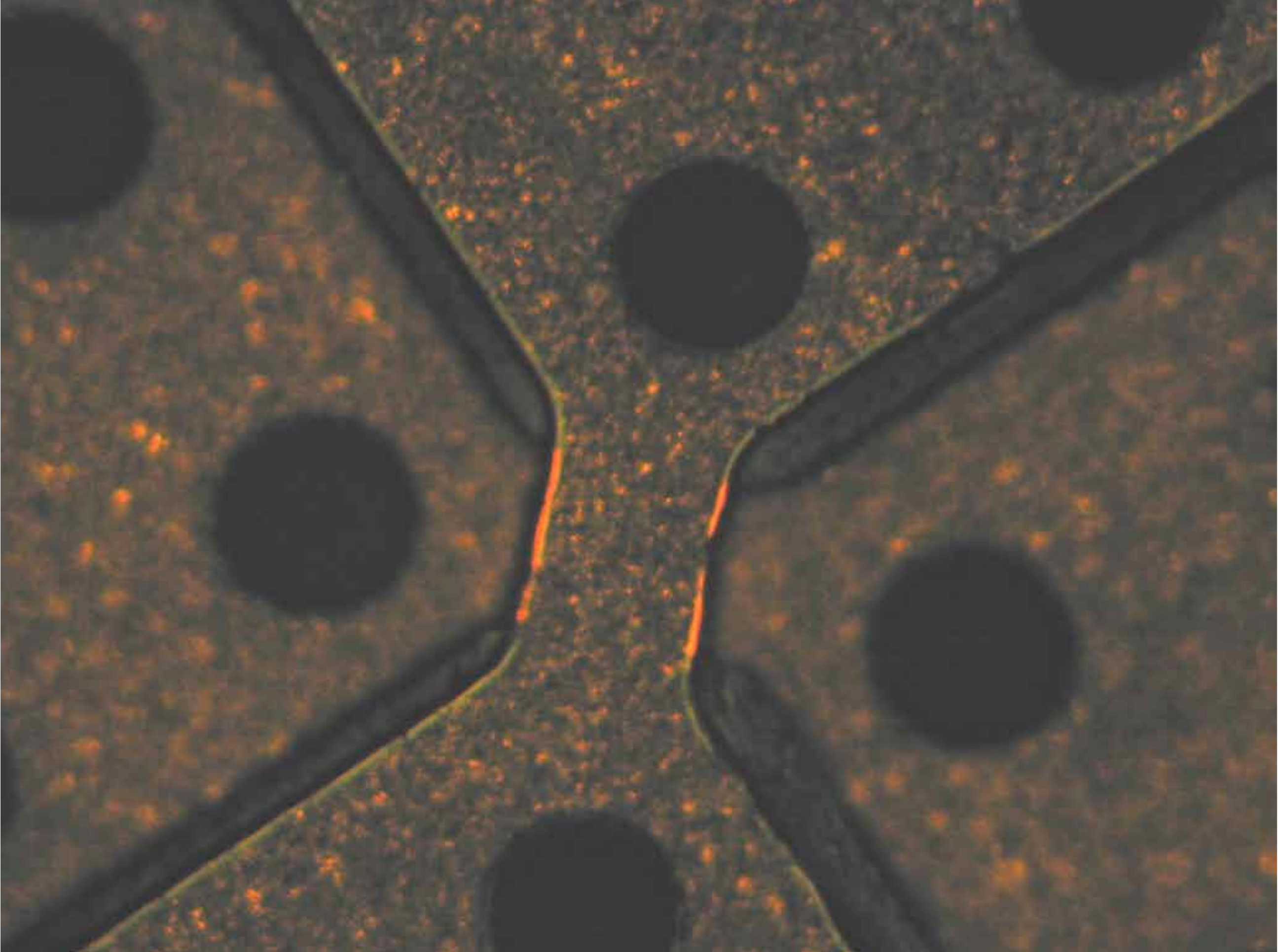} 
\end{tabular}
\end{center}
\caption{SILEM $x-y$ pixels design.}
\label{pixel}
\end{figure}

The SILEM grid consists of a main substrate made in a printed-board circuit, $400 \mu m$ thick,  with $18 \mu m$ thick copper films on both sides. The copper of the "bottom" side is patterned with a circular shape. The other side is patterned with square pixels, as shown on Figure \ref{pixel}. The SILEM pixels have $1mm$ sides. They are designed to have 4 holes in each of them. The pixels on the main substrate are linked in $x$ lines with $250 \mu m$ wide bridges on two opposite corners.
The $y$ pixel lines are patterned on a $20 \mu m$ thick kapton foil. This kapton foil is then cut off inbetween the $y$ pixel lines to open the space for the $x$ pixels. Finally, this $y$ plane is aligned and glued on the main substrate. Figure \ref{pixel} shows both $x$ and $y$ pixels on the upper side of the SILEM. The illustrated SILEM version (Fig \ref{pixel} left) has the $20\mu m$ kapton foil not opened. The kapton of the final SILEM version is opened.     

\begin{figure}[h]
\begin{center}
\begin{tabular}{cc} 
\includegraphics[scale=0.25]{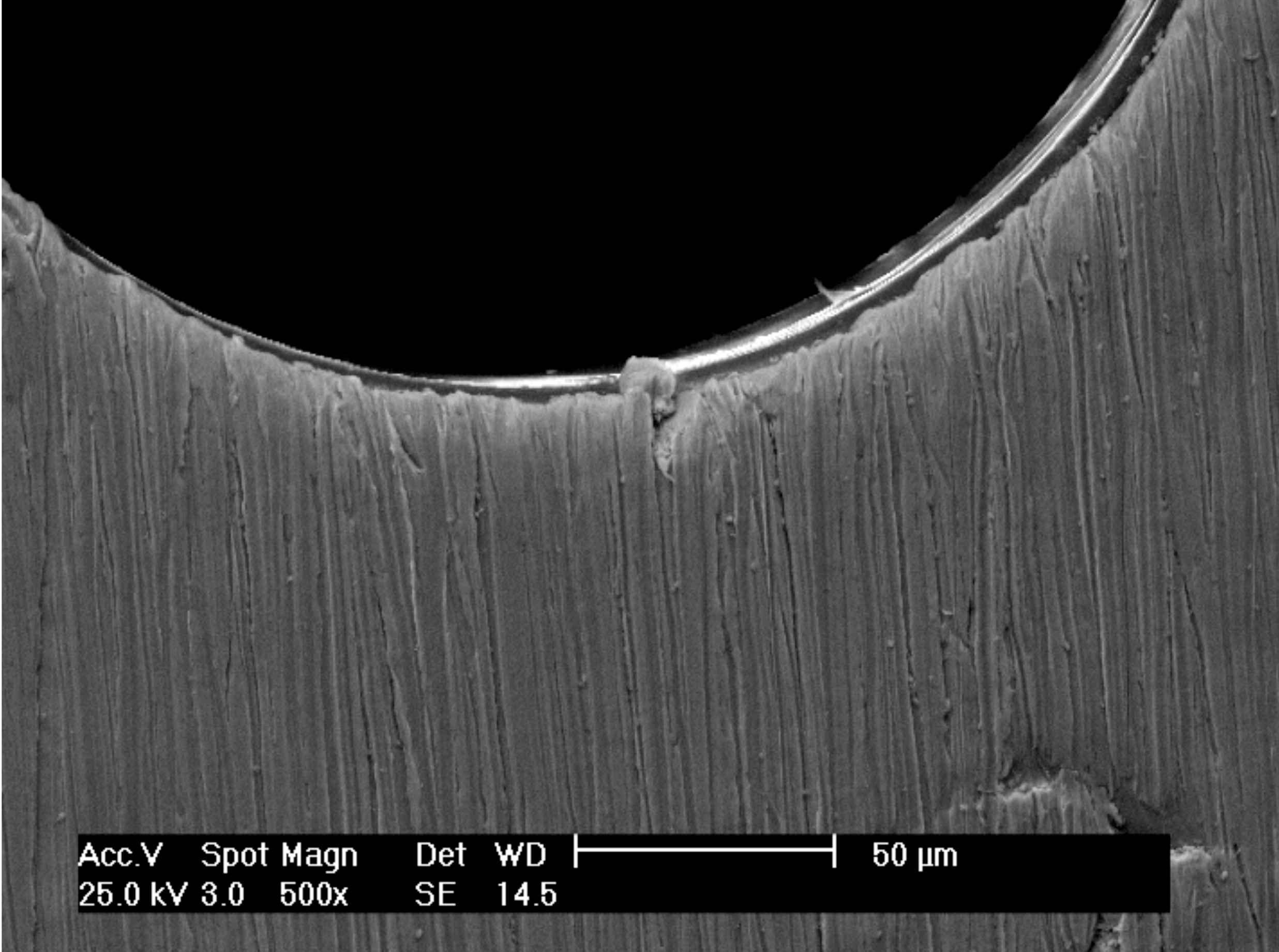} & 
\includegraphics[scale=0.25]{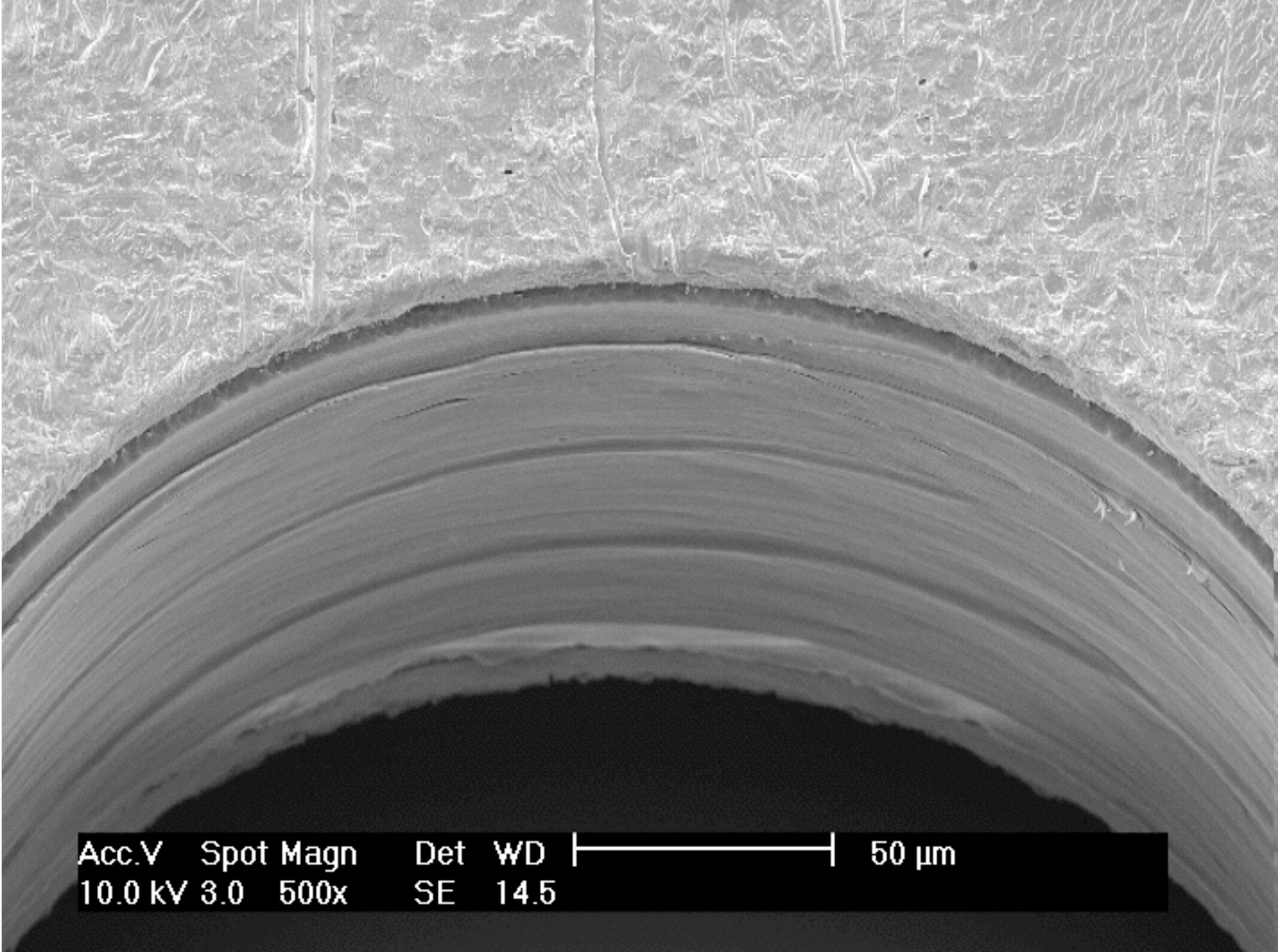} 
\end{tabular}
\end{center}
\caption{SILEM holes, before and after Cu etching.}
\label{etching}
\end{figure}

The $300 \mu m$ holes are then mechanically drilled. A chemical etching of the copper is necessary after the drilling process to remove the shavings around the holes, otherwise leading to breakdowns and sparks, due to sharp point effects. Figure \ref{etching} shows a SILEM hole, before and after the chemical etching of copper. The quality of the holes borders is clearly cleaner after this treatment. 

\section{Experimental results}

\begin{figure}[h!]
\begin{center}
\begin{tabular}{cc} 
\includegraphics[scale=0.35]{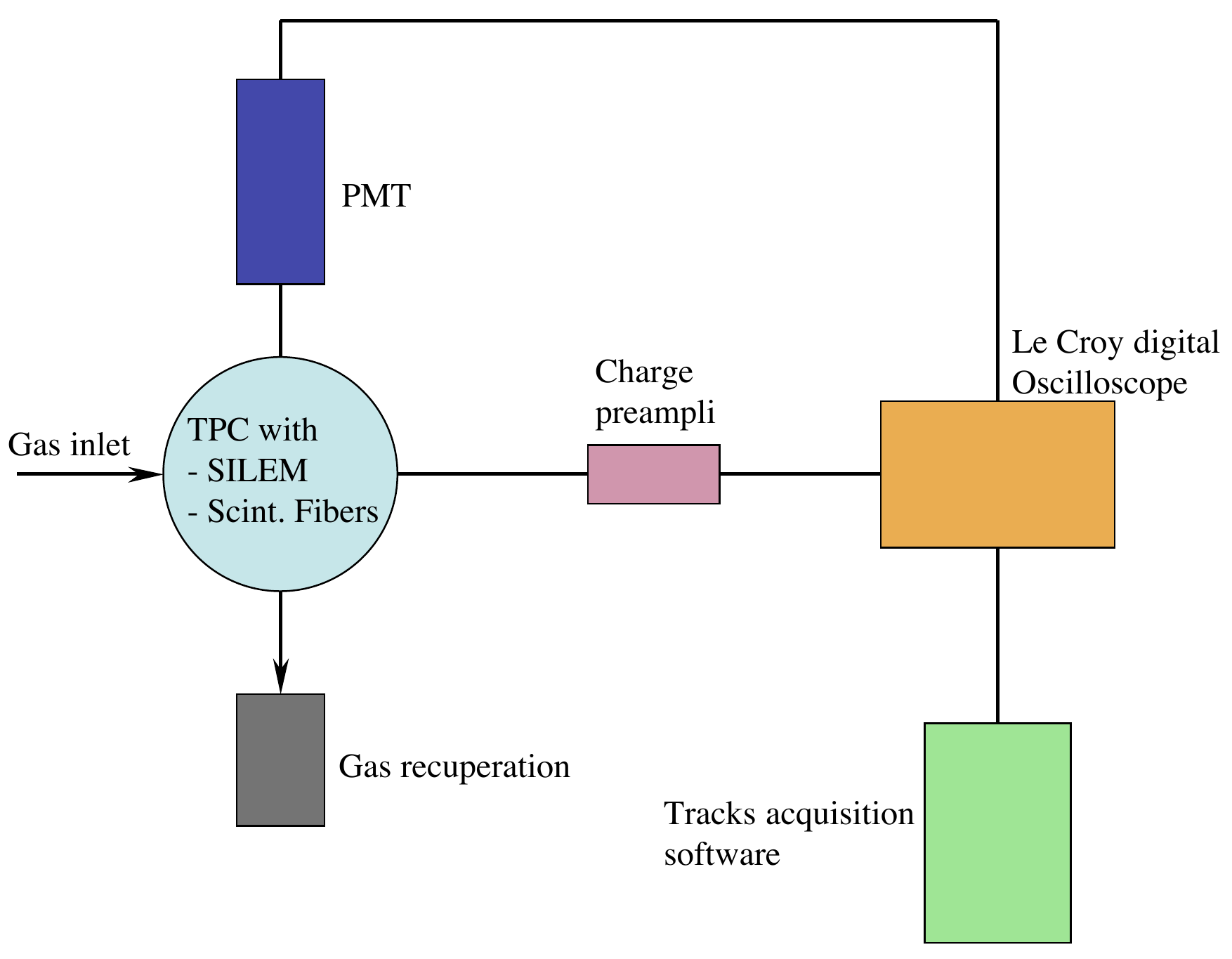} & 
\includegraphics[scale=0.26]{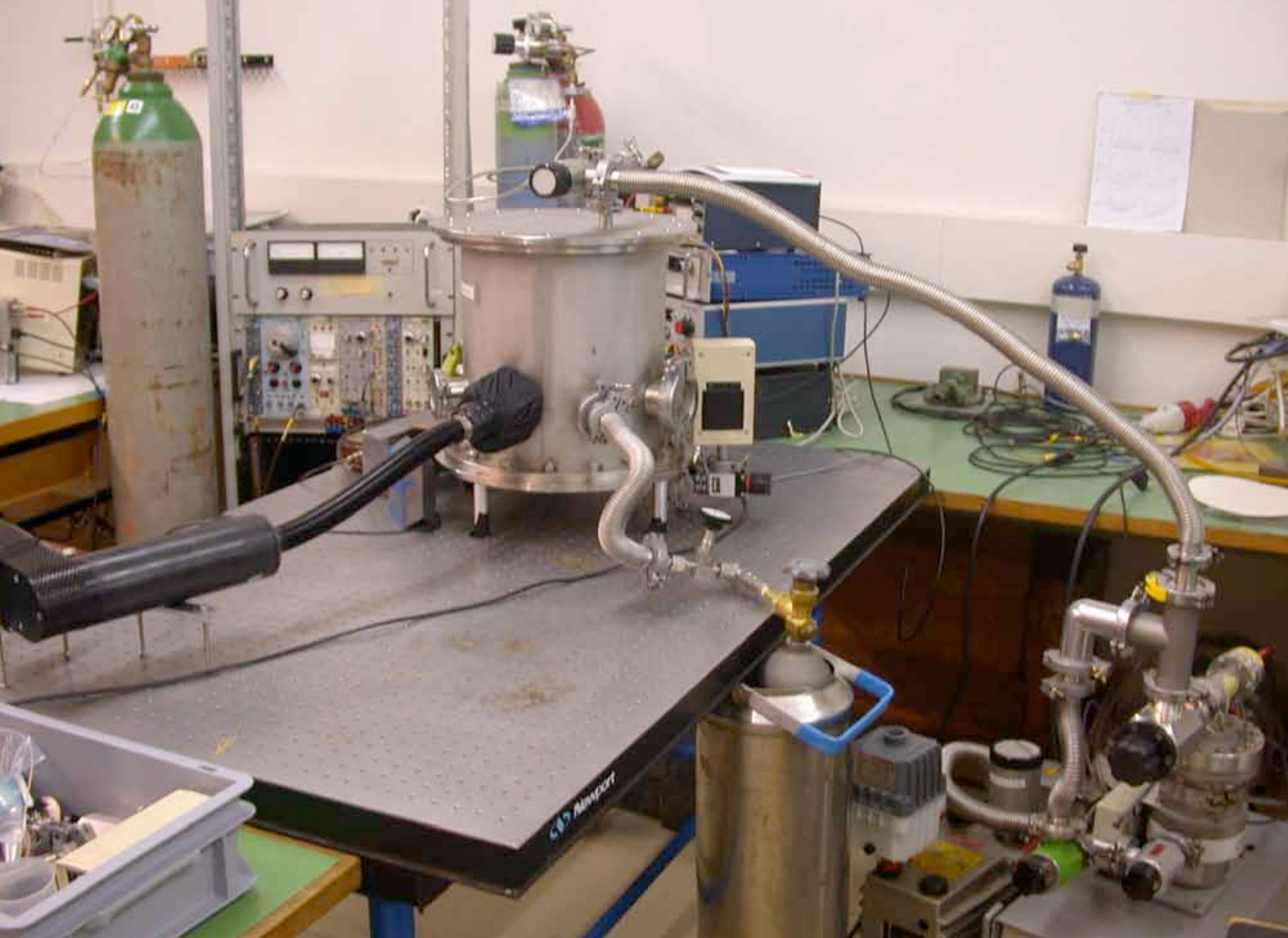} \\
\end{tabular}
\end{center}
\caption{Technical drawing of the experimental setup (left). Custom made TPC for SILEM measurements.}
\label{tpc}
\end{figure}

The SILEM amplification grids were tested in a custom made TPC with $ArCH_{4}$ at various pressures.  Figure \ref{tpc} shows the experimental setup and the TPC used for testing the SILEM. The advantage of having both $x$ and $y$ readout planes on the same SILEM side becomes clear here : in a TPC, the primary electrons are drifted towards the grid with an electric field. The TPC anode is kept to a negative potential and the "upper" SILEM side is kept to ground. The "bottom" side of the SILEM is at a positive potential. The amplified signal can be read on both SILEM sides. The secondary electrons signal is read on the "bottom" side. As it is at a given potential, the signal has to be decoupled from the potential. However, the ions signal, at the "upper" side, can be directly read as it is at ground.

\begin{figure}[h!]
\begin{center}
\begin{tabular}{cc} 
\includegraphics[scale=0.29]{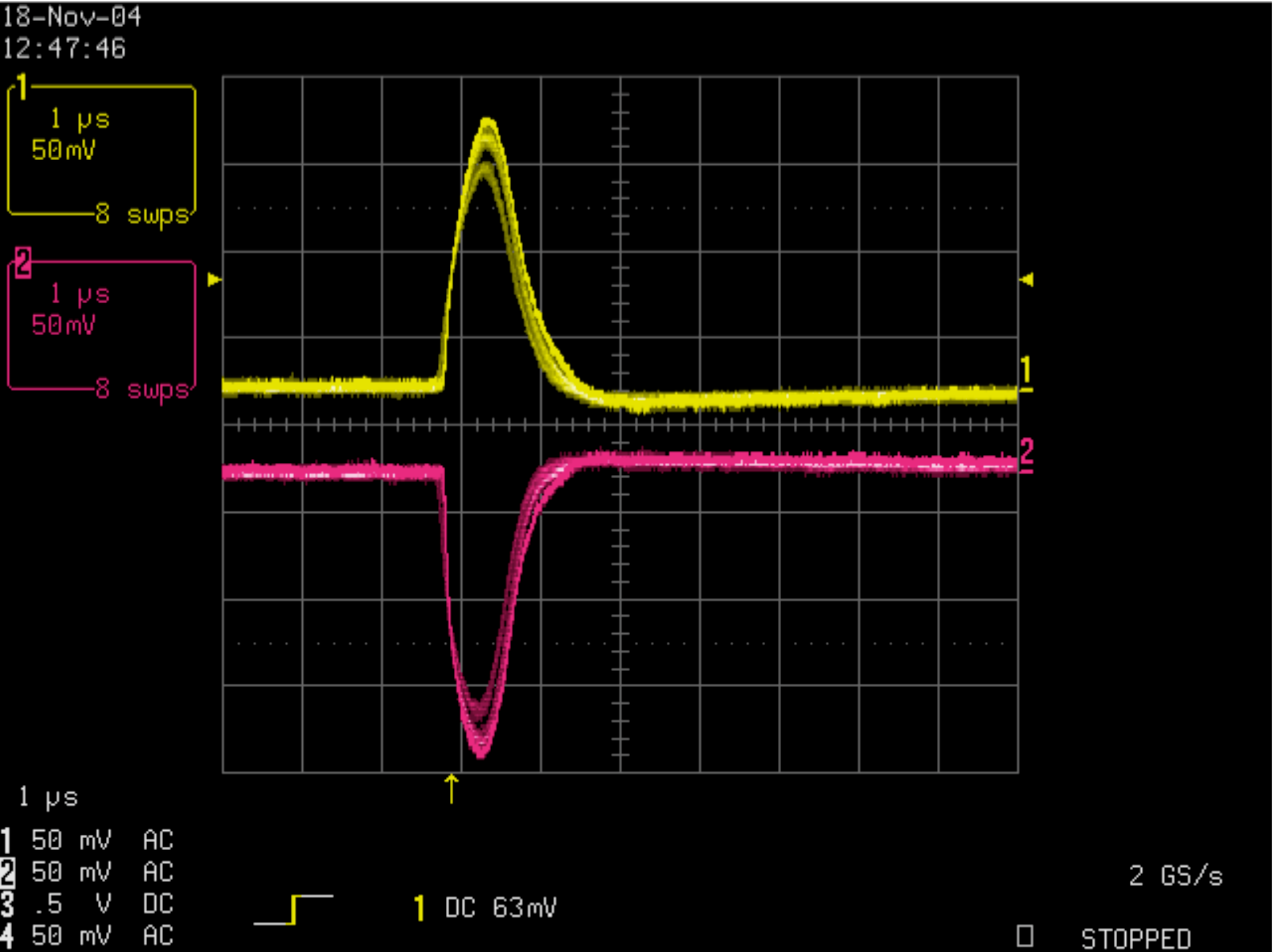} & 
\includegraphics[scale=0.4]{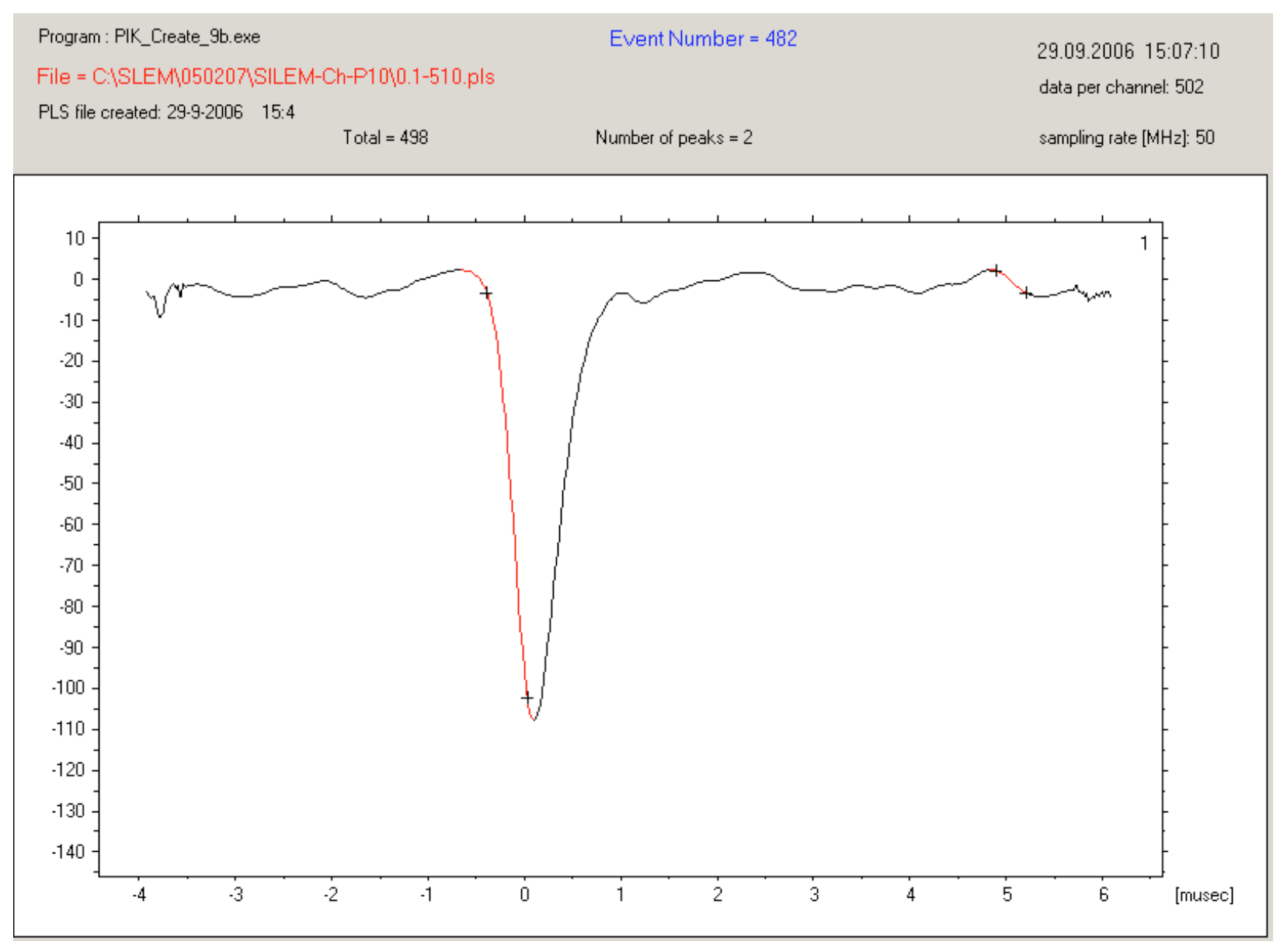} \\
\end{tabular}
\end{center}
\caption{SILEM charge signals.}
\label{charge}
\end{figure}

Figure \ref{charge} shows charge signals on both SILEM sides, i.e. ions (in yellow) and electrons (in red) charge signals. All signals are stored in a database. Their amplitude is proportional to the primary electron energy, i.e. proportional to the incident particles traversing the TPC volume. For the SILEM tests, we used a $^{55}Fe$ source emitting 5.9 keV x-rays. 

\begin{figure}[h!]
\begin{center}
\begin{tabular}{cc} 
\includegraphics[scale=0.4]{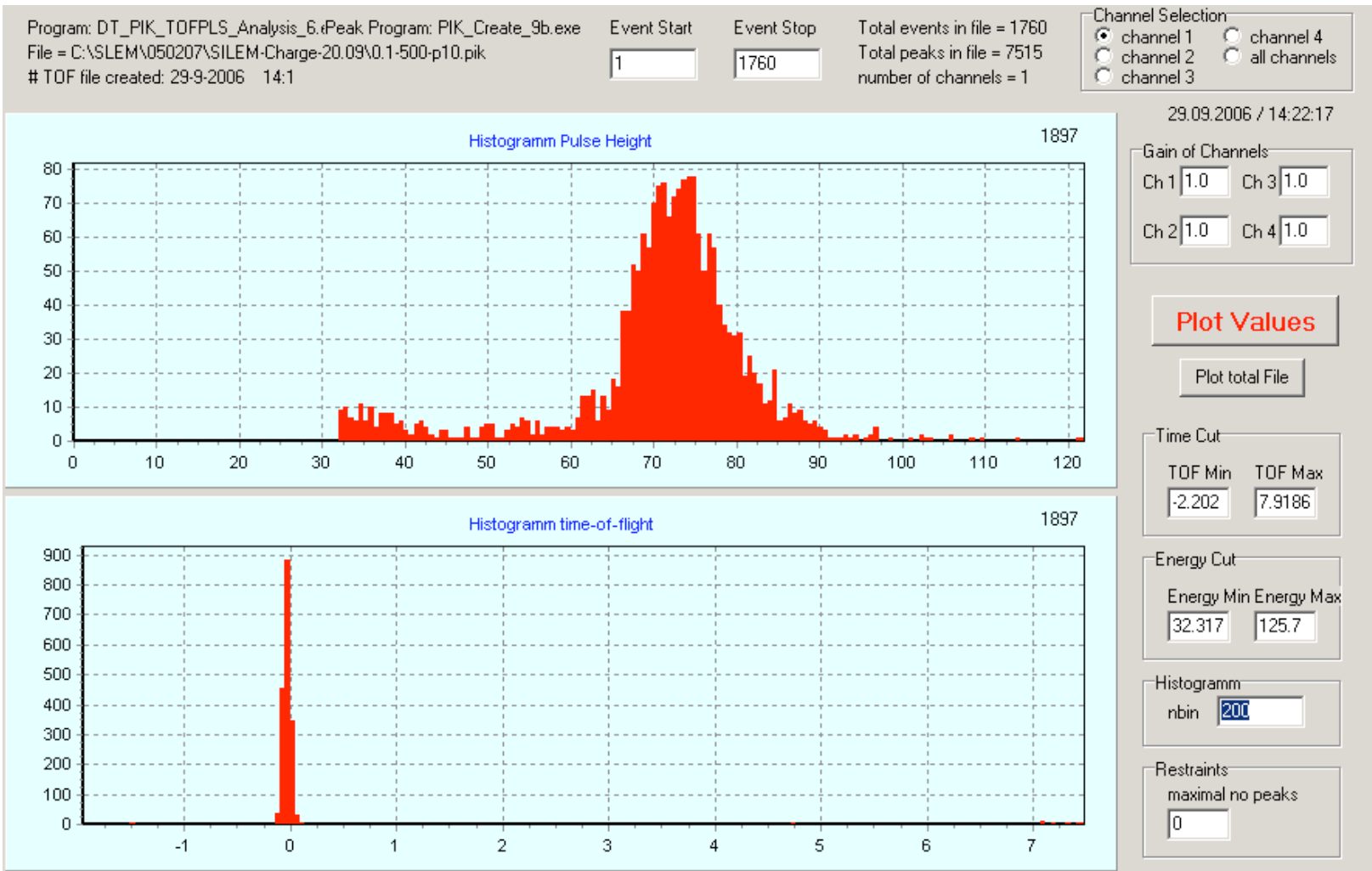} & 
\includegraphics[scale=0.36]{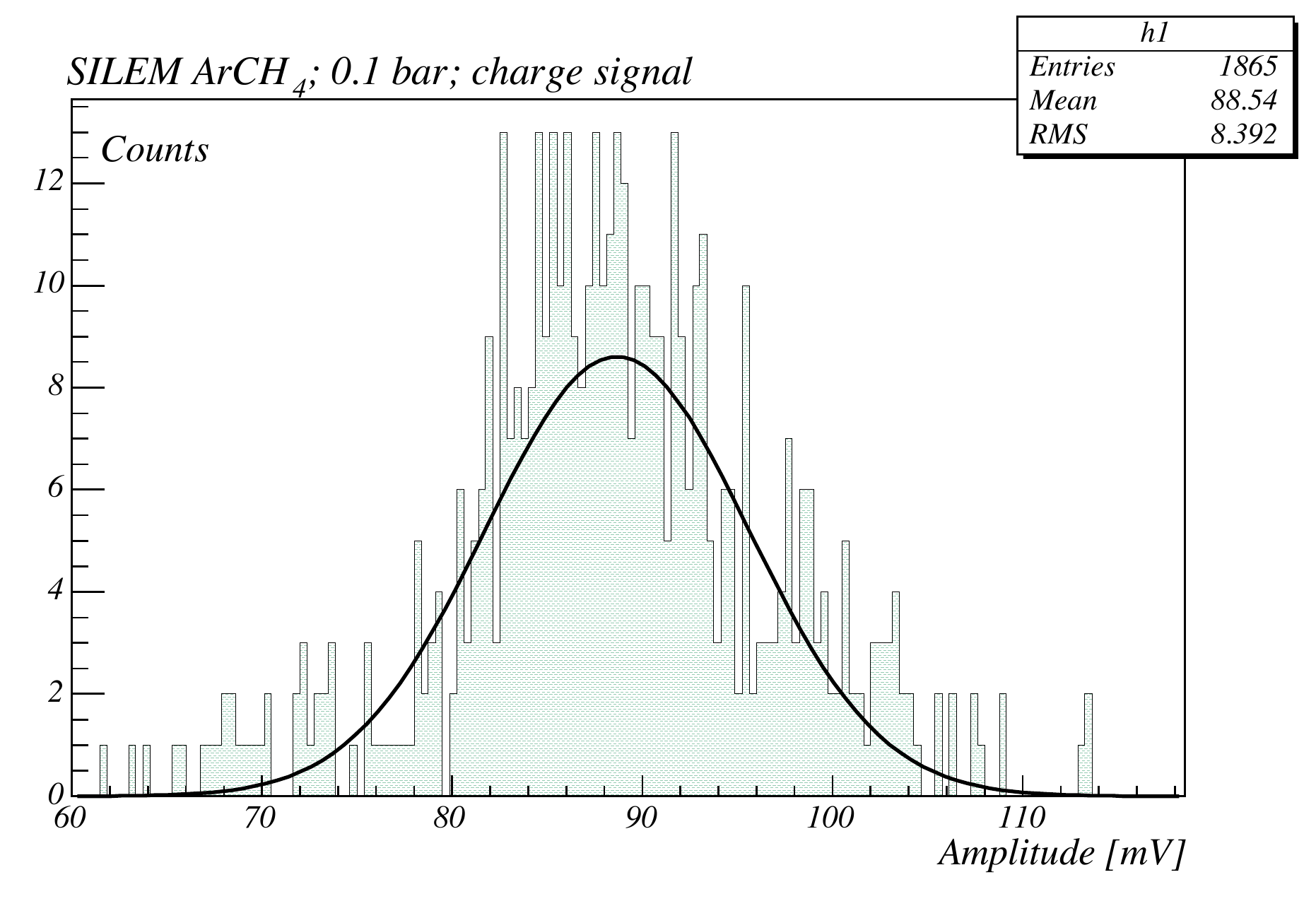} \\
\end{tabular}
\end{center}
\caption{SILEM measurements of the 5.9 keV $^{55}Fe$.}
\label{signal-treatment}
\end{figure}

Figure \ref{signal-treatment} shows the 5.9 keV peak of $^{55}Fe$. Measurements were performed in P10 with pressures in the range 100 mbar --- 2.0 bar.

\subsection{SILEM gain}

\begin{figure} [h!]
\begin{centering}
\includegraphics[scale=0.45]{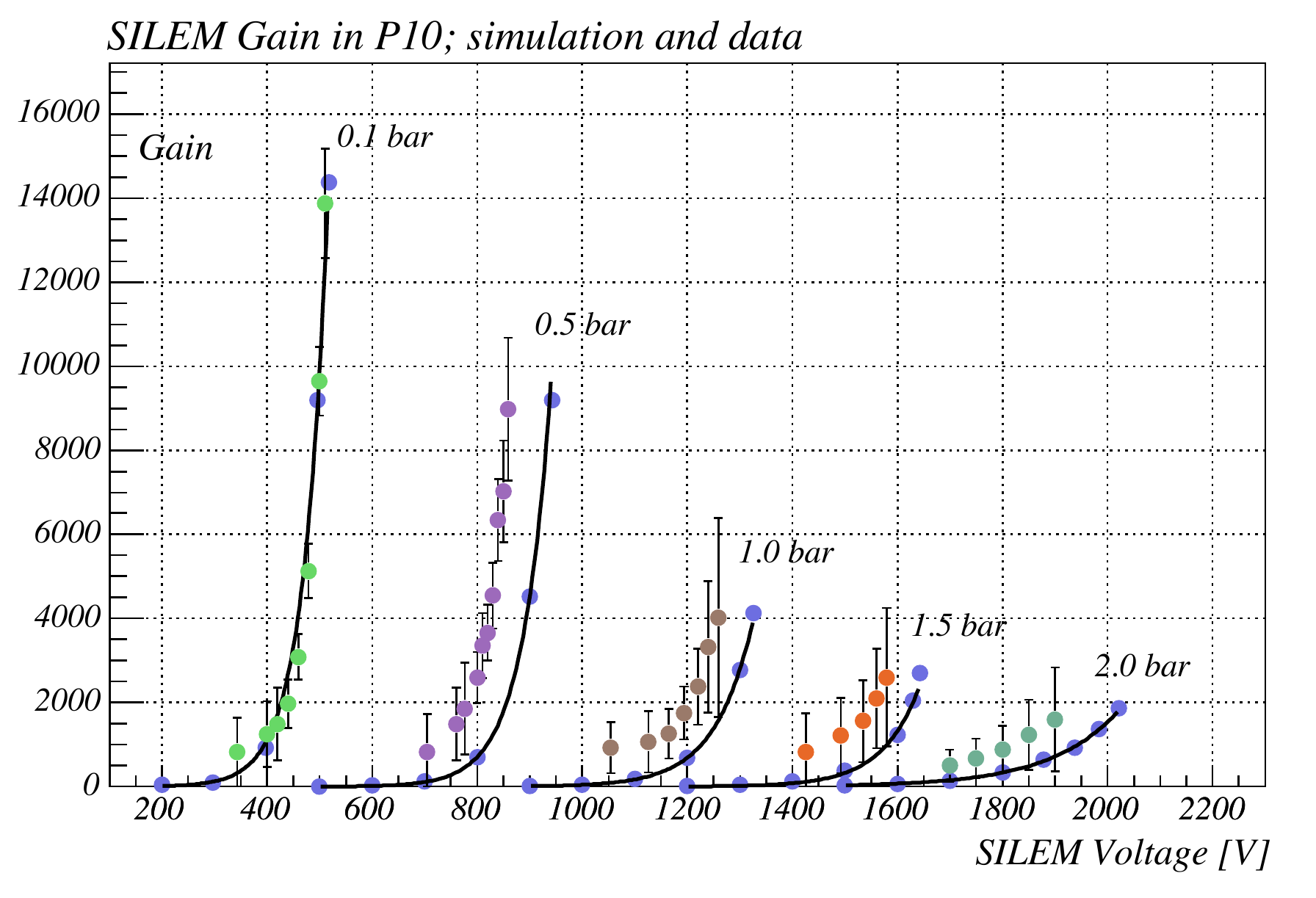}
\caption{SILEM gains in P10 at various pressures.}
\label{SILEM-gain}
\end{centering}
\end{figure}

The SILEM gain $G$ is proportional to the signal amplitude according to the relation :
$$G=\frac{C \ w}{e \ E} \ U$$
where $C$ is the preamplifier input capacitance, $w$ the ionization energy of the gas, $E$ the x-ray energy and $U$ is the signal amplitude before preamplification.\\
Figure \ref{SILEM-gain} shows the SILEM measured gains in P10, for pressures between 0.1 bar and 2.0 bar. The error bars correspond to the energy resolution of the $^{55}Fe$ peak. The fitted solid lines represents the Garfield simulation of the SILEM gain, which is in agreement with the experimental data.
The minimum of the experimental data points corresponds to the appearance of a charge signal from the  electronic background. The maximum corresponds to the beginning of sparks. 

\begin{figure} [h!]
\begin{centering}
\includegraphics[scale=0.45]{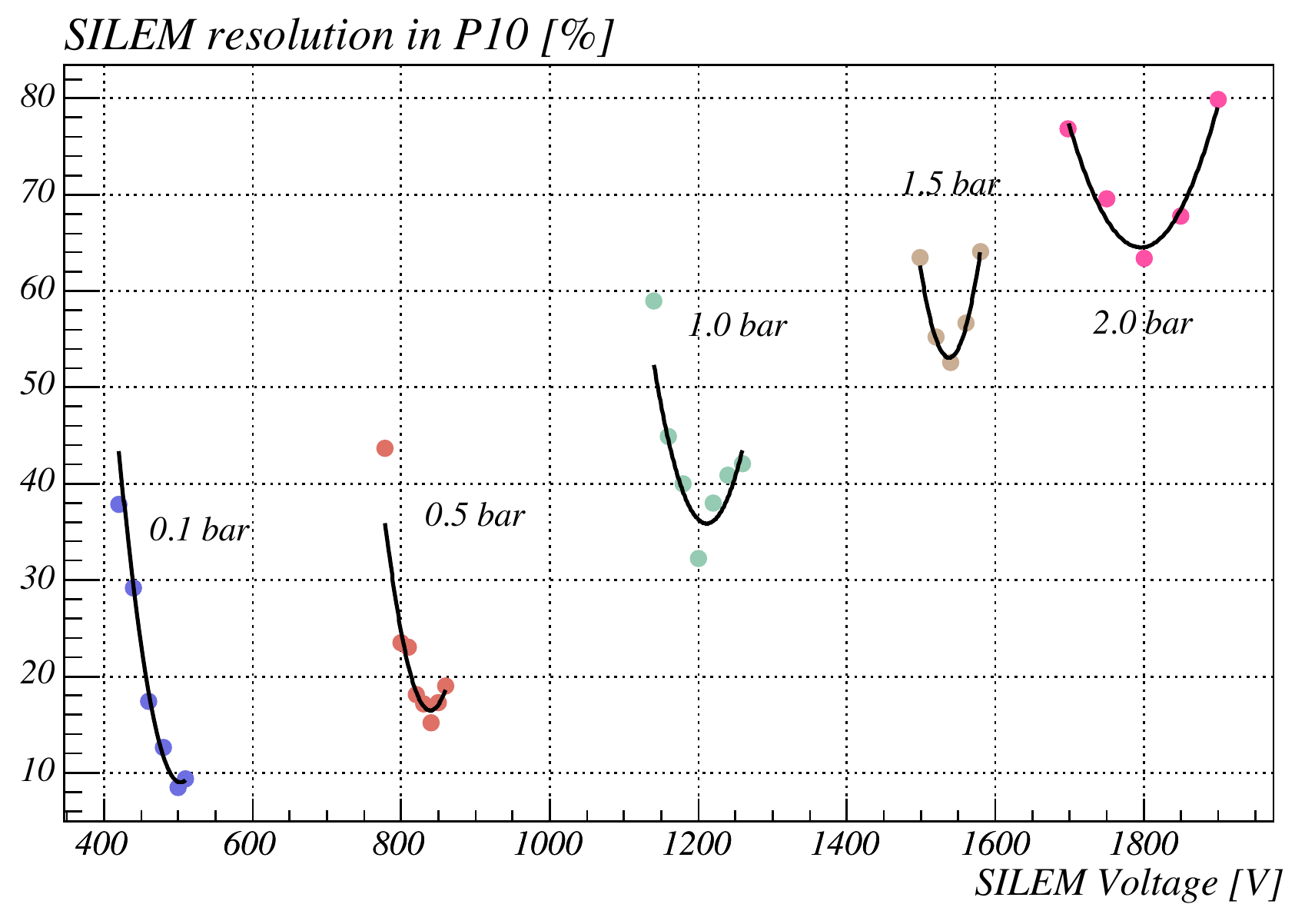}
\caption{SILEM resolutions in P10 at various pressures for 5.9 keV.}
\label{SILEM-resol}
\end{centering}
\end{figure}

As shown on Figure \ref{SILEM-resol}, the SILEM resolution increases with the P10 pressure. Moreover, for each pressure, there is a SILEM voltage corresponding to a minimal resolution.
These values were measured with a 5.9 keV $^{55}Fe$ source. A theoretical extrapolation of the resolution at 2 MeV would lead to resolutions smaller than $1\%$ even at 2.0 bar. Such resolutions would be useful for double-beta decay experiments for example.

\subsection{SILEM readout plane}

\begin{figure} [h!]
\begin{centering}
\includegraphics[scale=0.45]{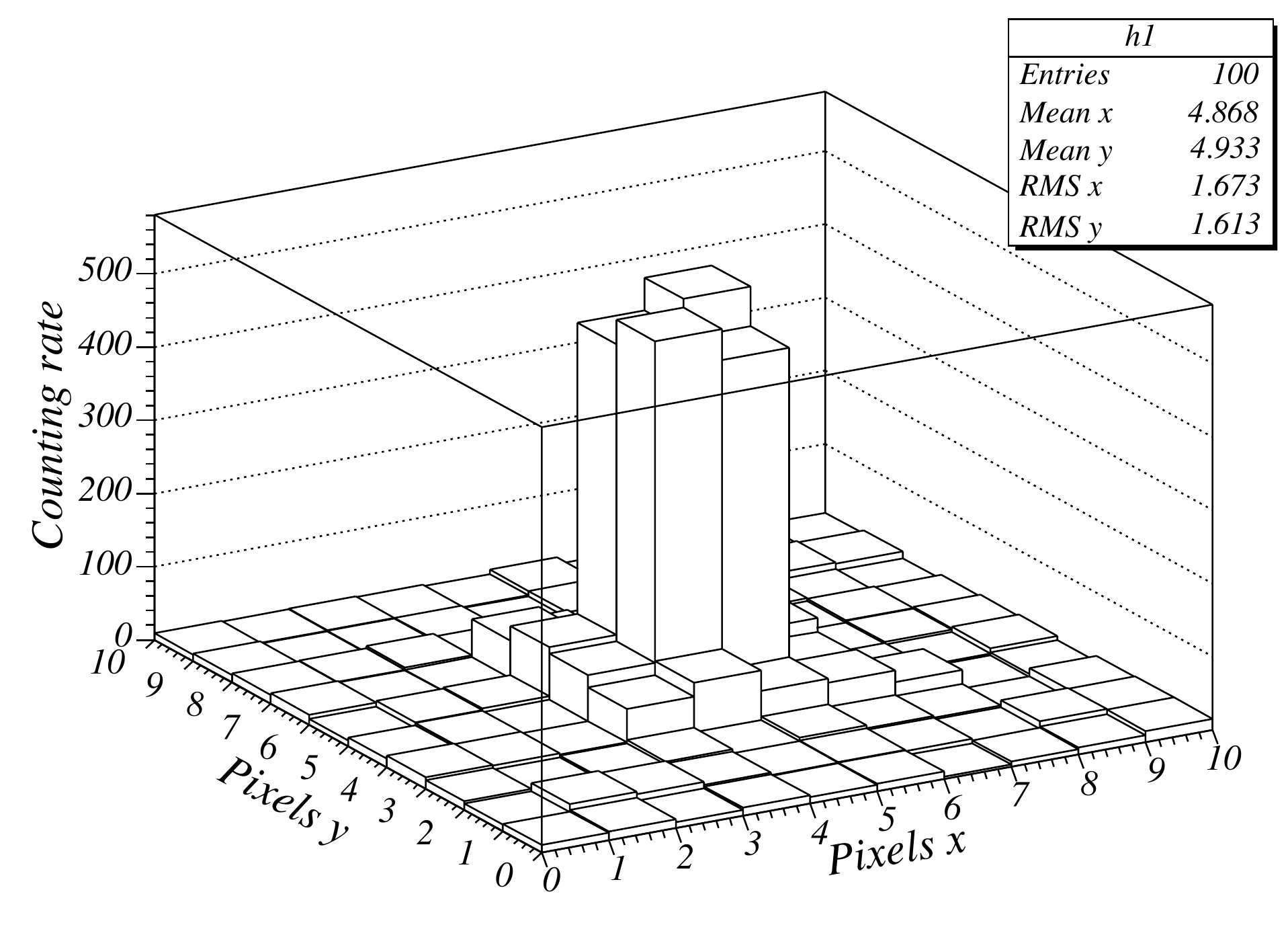}
\caption{SILEM pixels readout in $P10$ --- 1 bar.}
\label{pixels}
\end{centering}
\end{figure}

The $^{55}Fe$ source was collimated in such a way to irradiate only a $2 \times 2  \ mm^{2}$ surface on the SILEM grid; primary electrons are thus produced only in this limited surface. 10 SILEM channels were then scanned in the $x$ direction and 10 other in the $y$ direction. Figure \ref{pixels} shows that a significant signal has been measured only in the 4 pixels corresponding to the irradiated region of the grid. One can observe a small signal all around the 4 central pixels. These are certainly due to transverse diffusion of the primary electrons. This test was performed in P10 at 1 bar.

\section{Conclusion}

Development and testing of a new hole-type gaseous detectors with integrated micro-patterned readout plane and sparks resistant has been done successfully in this work. Optimization of the LEM geometry was the first phase of this project. A Maxwell/Garfield simulation has been carried out to investigate and determine the geometry leading to the best electron amplification, taking into account the experimental limitations, particularly concerning the mechanical hole drilling. The best LEM geometry must have a ratio LEM thickness $t$/holes diameter $d$ :

$$1\leq \frac{t}{d}\leq2$$ 

Integrating these design constraints, the SILEM has been fabricated and tested in P10 gas. It showed measurable gains up to 2.0 bar. The gain decreases with the gas pressure, but it would be possible to measure exploitable gains at higher pressures.
The energy resolution also increases with the gas pressure. For 5.9 keV x-rays, the resolutions were comprised between $6\%$ and $62\%$ for P10 pressures in the range 0.1 bar to 2.0 bar. A theoretical extrapolation of these energy resolutions at higher energies lead to smaller numbers. In the case of double-beta decay experiments like EXO, the gaseous detectors should be able to measure electrons up to $\sim$2.5 MeV ($0\nu\beta\beta$ decay in $^{136}Xe$ : $Q_{\beta\beta}=2457 keV$). One would thus have a SILEM energy resolution of $0.37\%$ at 5 bar and 2457 keV.
With such energy resolutions, it would be possible to exploit the parabola shape of the gain across the hole profile. One would then improve the spatial resolution of the grid of a factor $\sim10$, as the gain varies with $20\%$ across the hole diameter.\\

Moreover, the SILEM is equiped with an integrated $x-y$ micro-patterned pixels readout plane. It combines thus electron amplification and position measurements on the same object. It has been shown that it was possible to measure the single pixel through which the primary electrons are passing. As both $x$ and $y$ pixel lines are on the same grid side, they are both kept to ground and therefore do not require a signal decoupling. The pixels size has been fixed to 1mm, but this size could be decreased to a few hundreds of microns. It is actually limited by the LEM holes diameter. \\

The SILEM size is not limited by the holes drilling process, but only by the micro-patterning of the copper pixels structure. Standard photolithography techniques allow the patterning of structures up to $\sim60 \ cm$. Moreover, one could imagine that several succesive photolithography techniques could be processed. The alignement of the different steps remains feasible.





\end{document}